\documentstyle[prd,aps,floats]{revtex}

\begin{document}
\newcommand{\rhophi}{$\rho_{\phi}$}	
\preprint{astro-ph/9809272}
\draft

%
% Remove this and closure after abstract, plus preprint number,
% in electronic submission
%
\input epsf

\renewcommand{\topfraction}{0.99}
\twocolumn[\hsize\textwidth\columnwidth\hsize\csname 
@twocolumnfalse\endcsname

\title{A classification of scalar field potentials with cosmological
scaling solutions}
\author{Andrew R.~Liddle$^{1,2}$ and Robert J.~Scherrer$^{3,4}$}
\address{$^1$Astronomy Centre, University of Sussex, Falmer, Brighton BN1
9QJ,~~~U.~K.\\$^2$(Address from 1st Oct 1998: Astrophysics Group, The 
Blackett Laboratory, Imperial College, London SW7 2BZ.)} 
\address{$^3$Department of Physics and Department
of Astronomy, The Ohio State University, Columbus, OH 43210~~~U.~S.~A.\\
$^4$NASA/Fermilab Astrophysics Center, Fermi National
Accelerator Laboratory, Batavia, IL 60510~~~U.~S.~A.}
\date{\today} 
\maketitle
\begin{abstract}
An attractive method of obtaining an effective cosmological constant at the 
present epoch is through the potential energy of a scalar field. Considering 
models with a perfect fluid and a scalar field, we classify all potentials 
for which the scalar field energy density scales as a power-law of the scale 
factor when the perfect fluid density dominates. There are three 
possibilities. The first two are well known; the 
much-investigated exponential potentials have the scalar field mimicking the 
evolution of the perfect fluid, while for negative power-laws, introduced by 
Ratra and Peebles, the scalar field density grows relative to that of the 
fluid. The third possibility is a new one, where the potential is a positive 
power-law and the scalar field energy density decays relative to the perfect 
fluid. We provide a complete analysis of exact solutions and their stability 
properties, and investigate a range of possible cosmological applications.
\end{abstract}

\pacs{PACS numbers: 98.80.Cq \hspace*{6.2cm} Preprint~astro-ph/9809272}

\vskip2pc]

%%%%%%%%%%%%%%%%%%%%%%%%%%%%%%%%%%%%%%%%%%%%%%%%%%%%%%%%%%%%%%%%%%%%%%%%
\section{Introduction}

The evidence in favor of a cosmological constant, or something very
much like it, playing a significant dynamical role in our present
Universe is becoming overwhelming. Most prominent have been the recent
measurements of the apparent magnitude--redshift relationship using
Type Ia supernovae \cite{Sup}, but other factors such as the consistently low
measurements of the matter density, including the baryon fraction in
galaxy clusters \cite{WNEF}, have also been pointing in that direction. While 
many
of these latter measurements are insensitive to the presence of a 
cosmological constant, there is some observational
motivation for a flat Universe from the favored location of the first
acoustic peak of the microwave background anisotropies, and some
theoretical motivation from a desire to utilize the simplest models
of cosmological inflation as the source of density perturbations. In
combination, these favor a present cosmological constant (in units of
the critical density) of $\Omega_\Lambda \sim 0.7$.

Since a genuine cosmological constant requires extreme fine-tuning in
order to have only begun to dominate recently, it is extremely
tempting to model the cosmological constant as an effective one. As
the supernovae observations are requiring an accelerating Universe,
which is precisely the definition of inflation, the minimal approach
is to assume that the same mechanism drives inflation now as is
presumed in the early Universe, namely the potential energy of a
scalar field. Three possibilities present themselves. The field could
be at an absolute minimum of non-zero potential energy. It could be in
a metastable false vacuum, tunneling at some later stage into the true 
vacuum and perhaps even reheating. Or it could be slowly rolling down a
potential, as in the chaotic inflation models favored for the early
Universe.

The first two of these possibilities are dynamically indistinguishable
from a true cosmological constant, and so we shall concentrate on the
third, which is often called `quintessence'. As stressed in a recent
paper by Zlatev et al.~\cite{ZWS}, a rolling scalar
field offers an opportunity to address another mystery, that of why
the cosmological constant took so long to become dominant. If, for
example, the scalar field behaves in such a way as to remain
insignificant during the radiation domination era, perhaps it can be
`triggered' in some way to begin to grow in the matter era and come to
dominate only in the recent past. Solutions where the scalar field
energy density follows that of radiation or matter have been called
``scaling solutions", and more recently ``trackers", and several
examples have been described in the literature.

For the purposes of this paper, we will define a ``scaling solution"
as one in which the scalar field energy density, $\rho_\phi$
scales exactly as a power of the scale factor, $\rho_\phi \propto
R^{-n}$ when the dominant component has an energy density
which scales as a (possibly different) power:  $\rho \propto
R^{-m}$. An equivalent, and perhaps more fundamental, definition is that the 
scalar field kinetic and potential energies maintain a fixed ratio.
We will use the term ``tracker solution" to refer
to the special case $m=n$, i.e., where the scalar field
energy density scales in the same way as the dominant component.
The case where $m = n$ is produced by an exponential potential 
\cite{PLI,Wetterich,texas,FJ,CLW}, while negative
power-law potentials give $n < m$ \cite{ZWS,RP}.

In this paper we provide a comprehensive classification of all solutions of
this type, when the energy density is dominated by the perfect fluid.  We 
show that the {\em only} potentials which lead to this sort of
behavior are the previously studied exponential and negative power-law
potentials, and a new class of positive power-law potentials.  We study the
general properties of such solutions, including stability, and examine how
well they might do in giving the desired cosmological behavior.

\section{Scaling solutions}

A spatially-flat homogeneous Universe containing a perfect fluid with
energy density $\rho$ and pressure $p$, plus a scalar field $\phi$
with potential energy $V(\phi)$, satisfies the equations
\begin{eqnarray}
H^2 & = & \frac{8\pi G}{3} \left[ V(\phi) + \frac{1}{2} \dot{\phi}^2 +
	\rho \right] \,; \\
\dot{\rho} & = & -3H(\rho + p) \,,
\end{eqnarray}
where dots are time derivatives. We will assume that the perfect fluid has 
equation of state $p = (\gamma - 1)\rho$, which immediately implies
\begin{equation}
\rho \propto \frac{1}{R^m} \quad ; \quad m = 3\gamma \,.
\end{equation}
The scalar field $\phi$ evolves according to
\begin{equation}
\label{phiev}
\ddot{\phi}  =  - 3 H \dot{\phi} - \frac{dV}{d\phi} \,.
\end{equation}
The total scalar field energy is
\begin{equation}
\label{rhosum}
\rho_\phi = V(\phi) + \frac{1}{2} \, \dot{\phi}^2 \,,
\end{equation}
and we are interested in solutions for which $\rho_\phi \propto R^{-n}$
when $\rho_\phi \ll \rho$ and $\rho \propto R^{-m}$.

Eq.~(\ref{rhosum}) allows the scalar field equation to be written in the 
useful form
\begin{equation}
\label{rhophi}
\dot{\rho}_\phi = -3 H \dot{\phi}^2 \,.
\end{equation}

If we divide Eq.~(\ref{rhophi}) by $\rho_\phi$ and use
$\dot \rho_\phi / \rho_\phi = - n (\dot R /R)$, then we obtain
\begin{equation}
\label{fraction}
\frac{\dot{\phi}^2/2}{\rho_\phi} = \frac{n}{6} \,.
\end{equation}
Thus, power-law behavior for the scalar field energy density
requires that the scalar field kinetic energy
remain a fixed fraction of the total scalar field energy. The converse is 
true as well. This
makes sense: if the
kinetic energy evolves to become either dominant or negligible, then
$\rho_\phi$ will scale as $1/a^6$ or remain constant,
respectively. The former is not what we want, and the latter no
different from a genuine cosmological constant.
These two extreme cases also delimit the possible scaling
behavior for the scalar field energy density:
$0 \le n \le 6$, with the lower limit corresponding to
potential energy domination and the upper limit to kinetic-energy
domination.

\subsection{Exact Solutions}

Our basic method of solution is to assume the desired behavior of $\rho_\phi$
and $\rho$ and substitute into Eq.~(\ref{phiev}).  A similar procedure was
first undertaken by Ratra and Peebles \cite{RP}, who confined their attention
to the cases of matter and radiation domination, $m = 3,4$, and were
interested in certain classes of solutions.  Our development parallels and
extends their analysis.

When the perfect fluid with $\rho \propto R^{-m}$ dominates,
\begin{equation}
R \propto t^{2/m} \,,
\end{equation}
and Eq.~(\ref{phiev}) becomes
\begin{equation}
\label{phiev2}
\ddot{\phi}  =  - {6 \over m} {1 \over t} \dot{\phi} - \frac{dV}{d\phi} \,.
\end{equation}
The desired scaling behavior for $\rho_\phi$, substituted
into Eq.~(\ref{fraction}), gives
\begin{equation}
\label{dotphi}
\dot{\phi} \propto t^{-n/m} \,.
\end{equation}

Consider first the case $m = n$.  Then Eq.~(\ref{dotphi})
can be integrated to give $\phi \propto \ln(t)$.
Substituting this into Eq.~(\ref{phiev2}) and solving
for $V(\phi)$, we obtain
\begin{equation}
V(\phi) = {2 \over \lambda^2} \left({6 \over m} - 1 \right) 
	\exp \left( -\lambda \phi \right)
\end{equation}
This is the well-investigated exponential potential \cite{PLI},
for the limiting case where $\rho_\phi \ll \rho$. Although $\lambda$ can be 
positive or negative, those cases are physically identical, simply 
corresponding to a reflection of the $\phi$ trajectory about the vertical 
axis.

Provided $\lambda^2 > m$, the unique late-time attractor is a
tracker solution with $\rho_\phi = (m/\lambda^2) \rho_{{\rm total}}$ 
\cite{Wetterich,RP}. For
example, the scalar field will redshift as $1/a^4$ during radiation
domination, and then switch to $1/a^3$ once matter domination
commences. Although we derived it assuming $\rho_\phi \ll \rho$, in fact this 
solution exists for any fractional scalar field density
$\Omega_\phi$, through the appropriate choice of $\lambda$.

While mathematically intriguing, such solutions seem uninteresting as
candidates for a cosmological constant.  During nucleosynthesis they behave
as radiation and hence act like extra neutrino species, and are limited to
$\Omega_\phi < 0.2$ during radiation domination and hence $\Omega_\phi <
0.15$ during matter domination, well below the desired density
\cite{texas,FJ}.  A similar constraint arises from suppression of density
perturbation growth \cite{FJ}.  Anyway, such a scalar field is presently
evolving like matter and so won't explain the supernova measurements even if
it were permitted with a more substantial density.

These bounds can be evaded if the field does not enter the scaling
regime until late in the cosmological evolution, e.g.~after
nucleosynthesis for the first bound, and after structure formation has been
initiated for the second. However, this requires that the scalar field
begin with more or less its present density, and so provides no
answer to the original fine-tuning problem.

Now consider the case $m \neq n$.  In this case, integrating
Eq.~(\ref{dotphi}) yields
\begin{equation}
\label{exact}
\phi = A t^{1-n/m} \,.
\end{equation}
The second integration constant has been absorbed by horizontal translation
of $\phi$. Substituting the required behavior into the scalar field
equation leads to the potential
\begin{equation}
V(\phi) = A^2 \left(1-\frac{n}{m} \right)^2 \left(\frac{6-n}{2n}\right)
	\left( \frac{\phi}{A} \right)^\alpha \,,
\end{equation}
where 
\begin{equation}
\label{alpha}
\alpha = \frac{2n}{n-m} \,.
\end{equation}
The constant of integration, which would otherwise appear in $V(\phi)$,
vanishes because for scaling we need the kinetic energy to be a
fraction $n/6$ of the scalar field energy density.

Scaling behavior can therefore occur provided the potential has a
power-law form. If the exponent $\alpha$ is negative, then $m > n$ and the
scalar field energy density grows compared to the matter, whereas if
it is positive the opposite is true.
We can rewrite Eq.~(\ref{alpha}) as
\begin{equation}
\label{nalpha}
n = \left({\alpha \over \alpha-2}\right) m.
\end{equation}
Since $m$ and $n$ are positive, Eq.~(\ref{nalpha})
shows that scaling solutions exist for positive $\alpha$
only when $\alpha > 2$ (in section III below, we consider
what happens for $\alpha \le 2$).

We have thus determined {\it all} potentials which give
power-law scaling of $\rho_\phi$ when the dominant
density component also scales as a power of $R$.  The negative
power-law and exponential potentials have been studied
in detail \cite{ZWS,RP,PLI,Wetterich,texas,FJ,CLW}; our new result is the 
existence of scaling solutions with the positive power-law potentials.

For most of these potentials, the differential equation
governing the evolution of $\phi$, Eq.~(\ref{phiev2}),
is nonlinear, and the solutions we have derived
for $\phi(t)$ are particular rather than general
solutions (in the study of nonlinear differential
equations, these are known as `singular solutions').  Hence, 
although there
can be no other potentials which produce scaling behavior,
there is as yet no guarantee that the potentials we have derived
produce general solutions (as opposed to singular solutions)
which display the desired scaling behavior.  Put another way,
we must show that the singular solutions we have derived
in this section are attractors of the equations of motion.

\subsection{Attractor structure}

The attractor structure of the exponential potential has
been analyzed in detail elsewhere \cite{CLW} so we
will not concern ourselves with that potential here.
The attractor structure of the negative power-law potentials
has been discussed by Ratra and Peebles \cite{RP} for the
cases $m = 3,4$.  We extend their discussion to the case
of arbitrary $m$, and also consider the case
of positive power laws.

We substitute a potential of the form $V(\phi) = V_0 \phi^\alpha$
into Eq.~(\ref{phiev2}).  However, note that the
multiplicative constant in front of $dV/d\phi$ can be absorbed
into a rescaling of $t$.  Henceforth, we assume such a rescaling
and write
\begin{equation}
\label{phiev3}
\ddot{\phi}  =  - {6 \over m} {1 \over t} \dot{\phi} - \phi^{\alpha-1}.
\end{equation}
For this rescaled equation, the constant $A$ in Eq.~(\ref{exact})
is
\begin{equation}
\label{Aeqn}
A = \left[\left({2 \over \alpha-2}\right)\left({6\over m} - {\alpha
\over \alpha - 2}\right) \right]^{1/(\alpha-2)}
\end{equation}
Note that $A$ is well-defined, and the solution given by Eq.~(\ref{exact})
valid, only for 
\begin{equation}
\frac{6}{m} - \frac{\alpha}{\alpha-2} = \frac{6-n}{m} > 0 \,,
\end{equation}
which is satisfied automatically as long as $n < 6$.

Following Ratra and Peebles \cite{RP}, we make the change of variables
\begin{equation}
t = e^{\tau} \quad ; \quad u(\tau) = 
	\frac{\phi(\tau)}{\phi_{{\rm e}}(\tau)}\,,
\end{equation}
where $\phi_{{\rm e}}(\tau)$ is the exact (singular) solution given
by Eqs.~(\ref{exact}) and (\ref{Aeqn}).  With these changes,
Eq.~(\ref{phiev3}) becomes
\begin{eqnarray}
u^{\prime \prime} & + & \left({4 \over 2-\alpha}+{6\over m} - 1 \right) 
	u^\prime + \\
 && \quad \quad \quad {2\over \alpha - 2} \left[ {\alpha \over \alpha -2} 
	- {6 \over m} \right] \left(u - u^{\alpha-1} \right) = 0 \,, 
\nonumber
\end{eqnarray}
where the prime denotes the derivative with respect to $\tau$.
This can be split into the autonomous system
\begin{eqnarray}
\label{autonomous}
u^\prime &=& p,\nonumber\\
p^\prime &=& \left( 1 - {6\over m} -{4 \over 2-\alpha}\right) p + \\
  & & \quad {2\over \alpha - 2}\left[ {6 \over m} - {\alpha \over \alpha -2}
  	\right] \left(u - u^{\alpha-1}\right) \,. \nonumber
\end{eqnarray}
For positive $\alpha$,
the interesting case is when $\alpha$ is an even integer, and then there are 
three critical points, all with $p=0$ and with $u = -1$, $0$ and $1$.
All three of these represent solutions which asymptotically approach
$\phi = 0$, $\dot \phi = 0$.  The $u=+1$ and $u=-1$ critical points,
when they are attractors,
represent solutions which asymptotically approach the singular solution.
They give mirror-image trajectories; the $+1$ attractor
represents solutions which go to $\phi=0$ from the positive $\phi$ direction,
while the $-1$ attractor gives solutions which approach $\phi=0$
from the negative $\phi$ direction.
The $u=0$ critical point corresponds to solutions in which
$\phi$ goes to zero faster than in the exact solution in Eq.~(\ref{exact}).

Linearizing Eqs.~(\ref{autonomous}) about the $u=1$, $p=0$ critical
point and solving for the eigenvalues $\lambda_{\pm}$ of small perturbations
about this point, we find
\begin{eqnarray}
\label{lambda1}
\lambda_{\pm} & = & {1 \over 2} - {3 \over m} - {2\over 2 - \alpha} \\
 && \quad \pm \sqrt{\left({1 \over 2} - {3 \over m} - {2\over 2 - \alpha}
 	\right)^2 + 2\left({\alpha \over \alpha-2} - {6 \over m}
 	\right)} \,. \nonumber
\end{eqnarray}
For the cases $m = 3$ and $m=4$, this equation reduces
to the Ratra--Peebles results \cite{RP}.
The behavior of these eigenvalues is somewhat clearer
if written in terms of $m$ and $n$, using Eq.~(\ref{alpha}):
\begin{equation}
\label{lambda2}
\lambda_{\pm} = \frac{2n - m - 6 \pm \sqrt{(2n - m - 6)^2 + 8m(n-6)}}{2m} \,.
\end{equation}
The necessary and sufficient condition for stability is
that the real part of both $\lambda_+$ and $\lambda_-$ be
negative.  If the quantity under the square root in Eq.~(\ref{lambda2}) is 
negative, then this corresponds to
the requirement that $2n - m - 6 < 0$ and gives a stable spiral.
Note, however, that because $n < 6$, the second
term under the square root is always negative.
Hence, if the quantity under the square root is positive
(so that both eigenvalues are real) then
$2n - m - 6 + \sqrt{(2n - m - 6)^2
+ 8m(n-6)} < 0$ whenever $2n - m - 6 < 0$.
Hence, the condition for stability is just $2n - m - 6 < 0$,
regardless of the value of the quantity under the square root
(although that will determine whether the stable singular
point is a stable spiral or a stable node).
In terms of $\alpha$, the stability condition is
\begin{eqnarray}
\alpha < 2 \left({6+m \over 6-m}\right) & \quad & 
	\mbox{Negative $\alpha$} \,; \\
\label{alphalow}
\alpha > 2 \left({6+m \over 6-m}\right) & & \mbox{Positive $\alpha$} \,.
\end{eqnarray}

The first of these is always satisfied, showing that the scaling solution for 
the Ratra--Peebles potentials ($\alpha < 0$)
is a stable attractor for all values of $\alpha$ (as noted by Ratra and
Peebles for $m = 3,4$).  For positive $\alpha$, however, the scaling solution
is a stable attractor only for sufficiently large $\alpha$.  For example, in
the matter-dominated era, attractor scaling solutions exist only for $\alpha
> 6$, while in the radiation-dominated era, this condition becomes $\alpha >
10$. For $m=3$, $\alpha=6$, and $m=4$, $\alpha=10$,
we have a vortex point at the singularity, which is 
neutrally stable but not an attractor.

\subsection{Phase plane analysis}

\begin{figure}[t!]
\centering 
\leavevmode\epsfysize=6.5cm \epsfbox{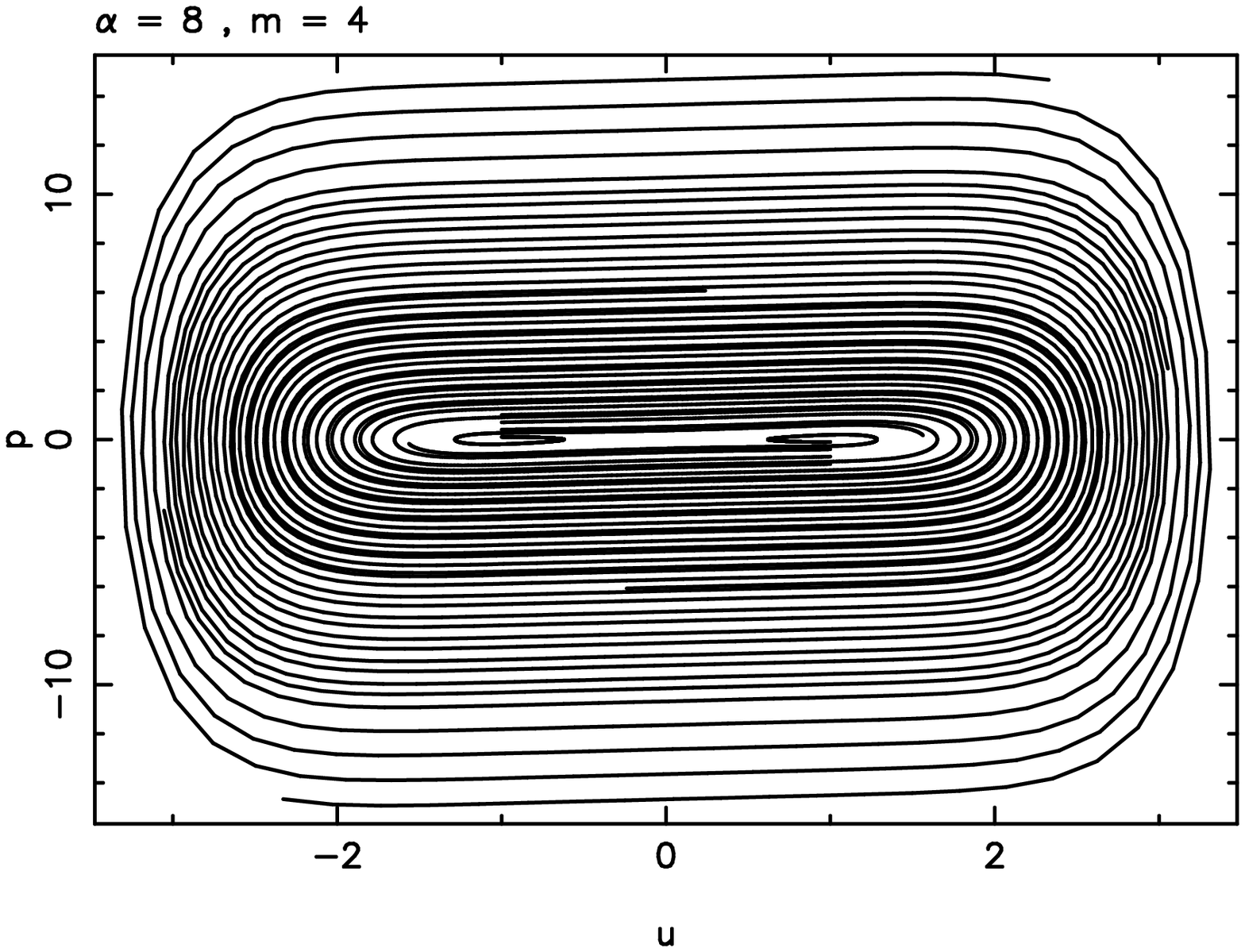}\\ 
\leavevmode\epsfysize=6.5cm \epsfbox{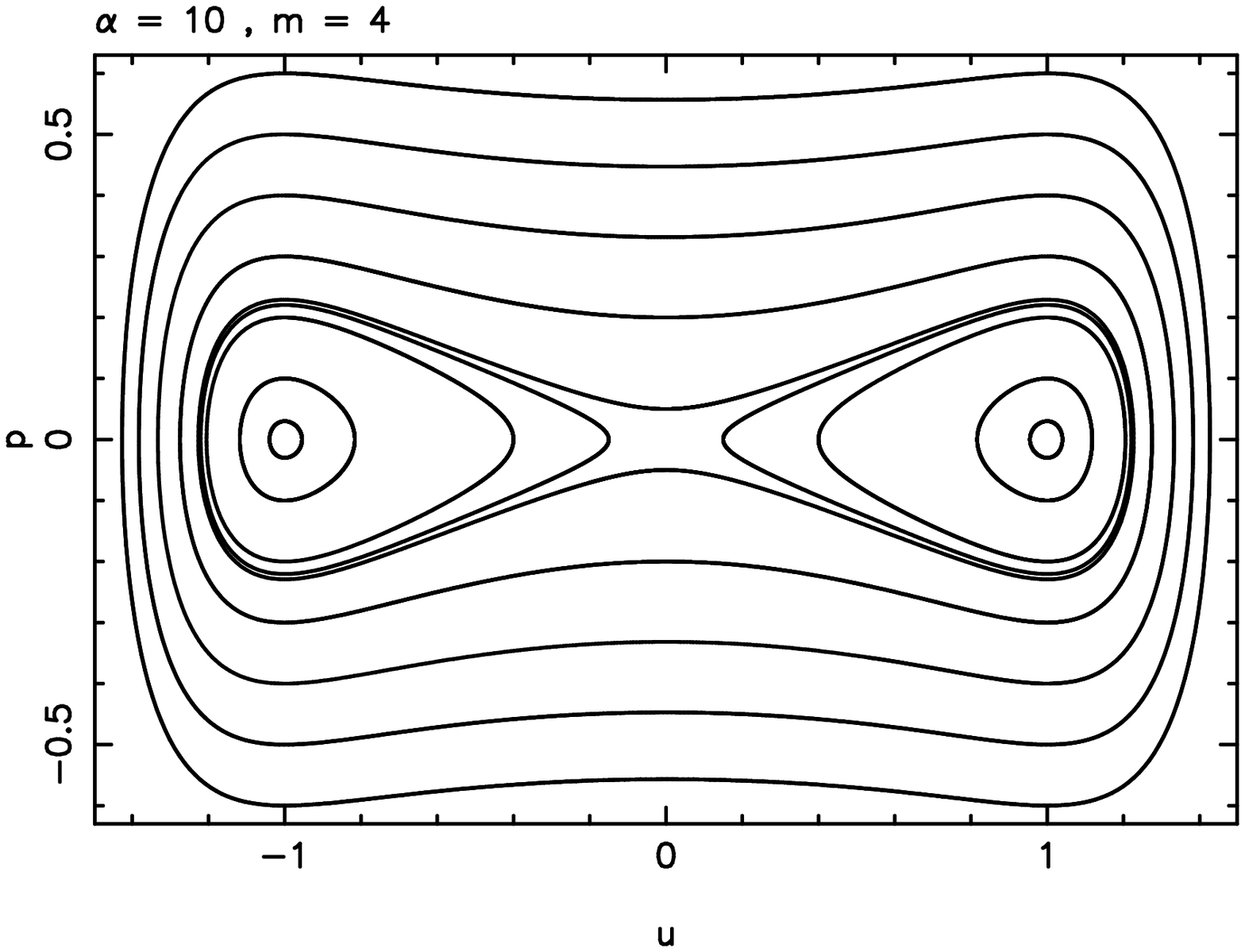}\\ 
\leavevmode\epsfysize=6.5cm \epsfbox{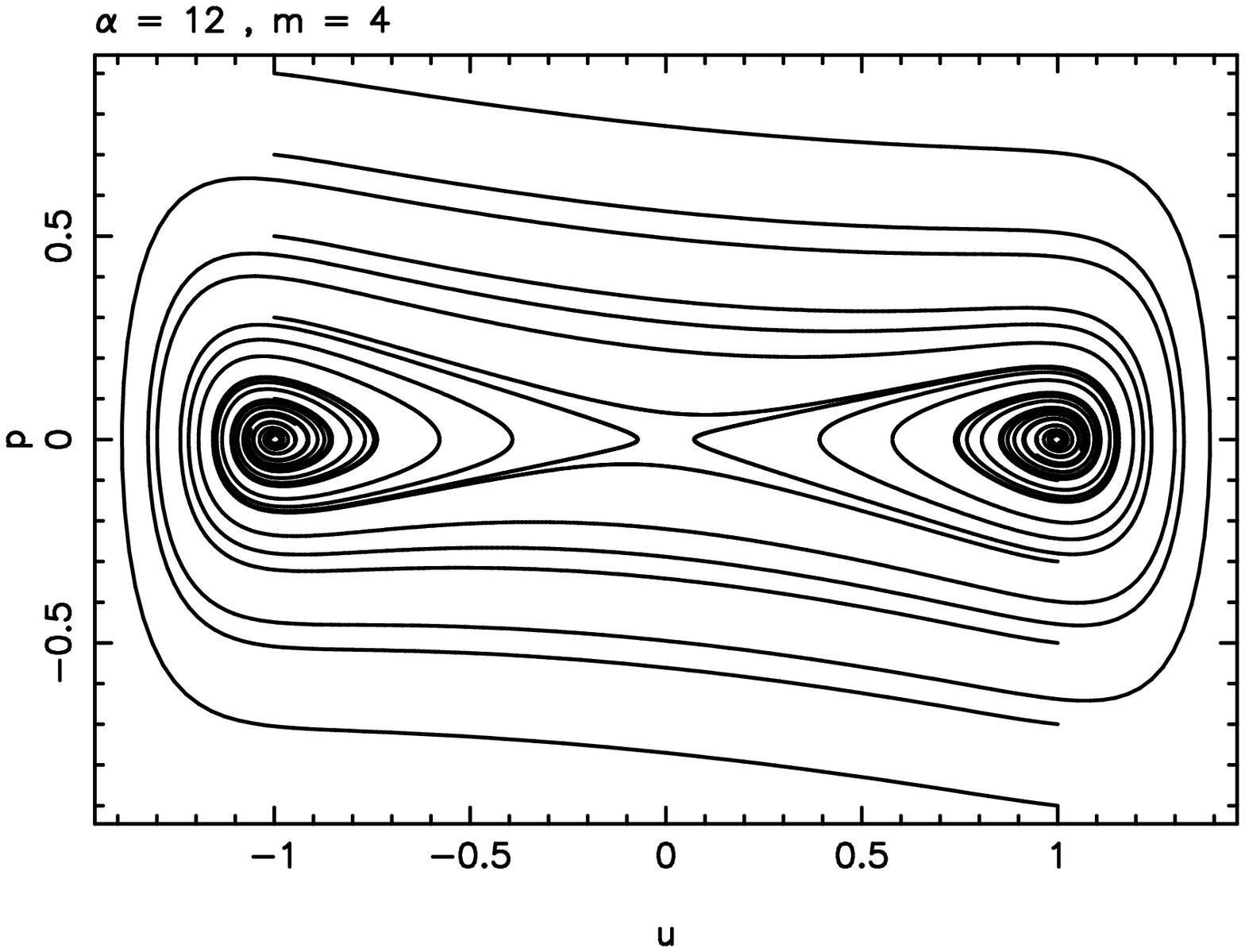}\\ 
\caption[phase_3]{\label{phase_3} Three phase planes for the 
radiation-dominated $m=4$ case. {}From top to bottom, $\alpha = 8$, $10$, 
$12$. The top panel spirals outwards, and the bottom one inwards.}
\end{figure}

A complete analysis requires a numerical solution.  Three phase planes are
shown in Fig.~\ref{phase_3} for the case $m=4$ (a radiation-dominated
universe).  {}From top to bottom, the choices are for the singular solution
[$u = \pm1$, $p=0$] to be unstable ($\alpha = 8$), to be marginally stable
($\alpha = 10$), and to be stable ($\alpha = 12$), respectively.  To
visualize the physical meaning of these trajectories, note that trajectories
which cross the $u=0$ vertical axis correspond to solutions in which the
$\phi$ field is oscillating about the minimum in the potential at $\phi=0$,
while trajectories which are confined to the left or right side of the $u=0$
axis correspond to solutions in which $\phi$ goes to zero without
oscillating.

In the top figure, any point in phase space spirals out to infinity.  This
corresponds to a solution which never stops oscillating. The amplitude of the 
$\phi$ oscillations is decreasing; the spirals move outwards because they 
lose amplitude more slowly than the exact solution $\phi_{{\rm e}}(\tau)$. 
Although such a solution exhibits a form of scaling behavior (see 
section~III.B below) the scaling
exponent is not given by Eq.~(\ref{nalpha}).  The middle figure
corresponds to marginal stability and the phase trajectories are closed loops
(which circulate clockwise), with the solution oscillating around the
singular solution but not approaching it.  The trajectories which cross the
$\phi = 0$ axis represent solutions in which $\phi$ oscillates forever.  The
trajectories closer to the singular point which do not cross the $\phi=0$
axis represent solutions in which $\phi$ does not oscillate about the
minimum, but approaches zero as $t^{1-n/m}$ times an oscillatory function.
The bottom figure shows the attractor situation; depending on the initial
conditions the trajectory may circulate several times around the two critical
points (which corresponds to the field oscillating about the minimum) before
circulating only about a single point (in which case the field stops
oscillating and falls steadily toward the minimum).  In the latter case the
exact solution given by Eq.~(\ref{exact}) is multiplied by an
oscillatory function which has a steadily decreasing amplitude.

We can repeat our stability analysis for the $u=0$, $p=0$ critical point.  We
find that the eigenvalues are both real, with $\lambda_- = (n-6)/m$, which is
negative for $n<6$, and $\lambda_+ = n/m - 1$, which is positive (since $n>m$
for positive $\alpha$).  Thus, the 
$(0,0)$ critical point is always an unstable saddle point.
This may seem bizarre, since the $(0,0)$ critical point corresponds to the
field lying motionless at the bottom of the potential.  However, remember
that the singular solutions also asymptotically reach the minimum, and our
result simply means that if the field is perturbed slightly from this
minimum, it returns to the minimum via the singular solution.

\section{Applications and special cases}

\subsection{Negative power-laws}

If $\alpha$ is negative we have a decaying power-law potential, in which the
field can roll forever.  These are the potentials first investigated by Ratra
and Peebles \cite{RP}.  They were recently reexamined in some detail by
Zlatev et al.~\cite{ZWS}, in the context of the current observational
situation.  Because the scalar field density grows relative to the fluid,
eventually the approximation that the fluid energy density is dominant will
break down.  When that happens, the Universe enters an inflationary regime,
which has in fact been investigated in the early Universe context under the
name ``intermediate inflation'' \cite{interinf}.  The expansion rate 
asymptotically becomes
\begin{equation}
R \propto \exp \left[ t^{4/(4-\alpha)} \right] \,,
\end{equation}
and the fluid becomes less and less relevant. The inflationary regime may be 
preceded by a period of non-inflationary scalar field domination, if the 
scalar field comes to dominate while $\phi$ is sufficiently small.

The scalar field density grows with respect to the fluid regardless of
whether the Universe is radiation or matter dominated, so these solutions do
not exhibit a `triggered' transition into the inflationary regime.  Rather,
the timing of that transition is governed by the initial conditions, and for
the domination to be a recent event, one has to arrange for the initial
scalar field density to be well below the radiation density.  The tuning is
not however as severe as with a pure cosmological constant, since the
redshifting of the scalar field may be quite similar to that of the fluid
\cite{ZWS}.  A particularly interesting case arises for $\alpha = -6$; such a
scalar field will scale as matter during the radiation-dominated era, and
then grow relative to matter, as $\rho_\phi \propto R^{-9/4}$, once the
matter-dominated era begins.  If the field is generated initially with
$\rho_\phi \approx \rho_{{\rm matter}} \ll \rho_{{\rm rad}}$, then it will
continue to evolve with $\rho_\phi \approx \rho_{{\rm matter}}$ until matter
domination.  The onset of matter domination then triggers a change in the
evolution of the scalar field energy density, and $\rho_\phi$ begins to
evolve in a manner close to a curvature density until it comes to dominate.

An interesting question is whether it might be possible to find inflationary
scenarios capable of providing suitable initial conditions.  A possible
objection to the above is that in standard cosmological scenarios the energy
density which today is in non-relativistic particles (especially the baryons)
starts out as highly relativistic, only later to change its equation of state
on cooling, rather than already existing as a trace amount in the early
Universe.  However, the scenario just outlined bears some similarity to
suggestions for creation of the cold dark matter at the end of the
inflationary epoch \cite{heavyCDM}.

\subsection{Positive power-laws}

Although positive power-law potentials are more commonly associated with 
driving an inflationary expansion, provided they satisfy 
Eq.~(\ref{alphalow}) then we have shown that they too permit stable scaling 
solutions. In this case $m > n$, and so the scalar field becomes
progressively less important as the evolution proceeds, better and
better justifying the neglect of the scalar field terms in the
Friedmann equation.  The scaling solution for $\phi$ goes
smoothly to zero as $t \rightarrow \infty$, without oscillations.

Note, however, that these potentials (for even $\alpha$) can also support 
oscillatory behavior, with \cite{Turner}
\begin{equation}
\label{mturner}
\rho_\phi \propto R^{-6\alpha/(\alpha+2)} \,.
\end{equation}
Despite the power-law behavior, these solutions are not encompassed in our
definition of scaling, as the scaling law arises only after averaging over
oscillations, while within each oscillation, energy is continually being
converted between potential and kinetic.  When the field is oscillating, and
Eq.~(\ref{mturner}) applies, the scaling of $\rho_\phi$ with $R$ is
independent of the equation of state of the dominant component of the
density; this differs from the attractor solution in which the scaling of
$\rho_\phi$ with $R$ depends on $m$.  Furthermore, such solutions apply to
oscillating fields even when the scalar field density itself is dominant.  In
our phase diagram, Fig.~\ref{phase_3}, this oscillating solution corresponds
to the regime in which the phase space trajectory winds around both
attractors.

Whether the oscillatory behavior perseveres depends on the stability 
condition of Eq.~(\ref{alphalow}), which if satisfied implies that the 
scaling behavior found in Section~2 is the attractor. The oscillating 
solution has an amplitude $\phi_{{\rm max}} \sim R^{-6/(\alpha+2)}$, which 
matches the redshift dependence of the singular scaling solution if the 
stability condition is saturated, leading to the closed loops seen in the 
middle panel of Fig.~\ref{phase_3}. 

\begin{figure}[t!]
\centering 
\leavevmode\epsfysize=8.5cm \epsfbox{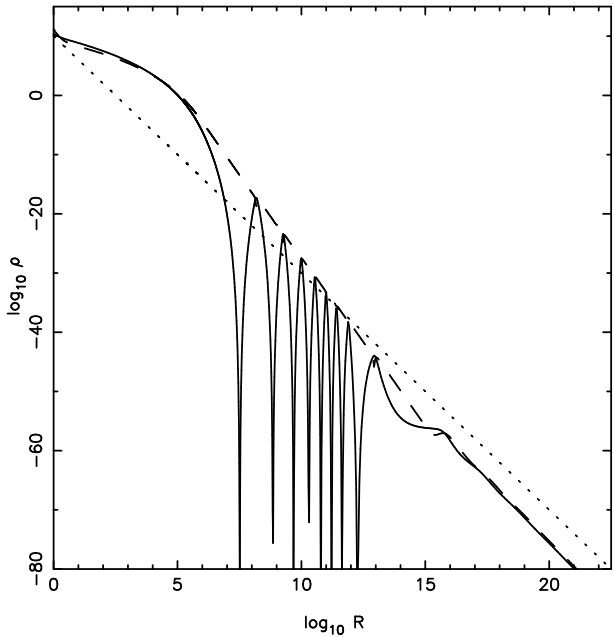}\\ 
\caption[alpha30]{\label{f:alpha30} The evolution of the energy densities in 
a radiation-dominated Universe with $\alpha = 30$. The vertical axis is 
in arbitrary units. The dotted line is the radiation energy density, and the 
solid line is the scalar field potential energy. The scalar field kinetic 
energy, shown as the dashed line, oscillates through zero out of phase with 
the potential energy, but does so too sharply for the plotting resolution.}
\end{figure}

If the stability condition is satisfied, this can lead to some interesting
behavior.  Consider the limit where $\alpha$ is very large and positive in
the radiation-dominated era.  If initially $\rho_\phi \gg \rho_{{\rm rad}}$,
but $\phi$ is oscillating rapidly, then the $\phi$ energy density will scale
roughly as $\rho_\phi \propto 1/R^6$ and eventually fall below the radiation
density.  When this happens, however, the scaling behavior will take over and
the $\phi$ energy density will scale as $1/R^{4+\epsilon}$ with $\epsilon \ll
1$.  We show this evolution with $\alpha = 30$ in Fig.~\ref{f:alpha30}.
Initially we have inflation, and then the field undergoes oscillations which
are heavily dominated by the kinetic energy, so its energy falls off at
nearly $1/R^6$, similar to kination \cite{kination}.  Finally the scalar
field becomes subdominant and stops oscillating, instead entering the scaling
solution with $\rho_\phi \propto 1/R^{(4 + 2/7)}$.  This scenario provides
yet another ``natural" mechanism to give a scaling solution with $\rho_\phi$
roughly equal to the density of the dominant component, since the scalar
field density drops rapidly relative to the radiation density until
$\rho_\phi \approx \rho_{{\rm rad}}$, after which $\rho_\phi$ decreases at
nearly the same rate as $\rho_{{\rm rad}}$.  A similar behavior has also been
noted for the case of negative power-laws when the $\phi$ density is
initially much larger than its attractor value \cite{ZWS}; the difference in
the case discussed here is that the attractor is reached even when $\rho_\phi
\gg \rho_{{\rm rad}}$ initially.

If the stability condition Eq.~(\ref{alphalow}) is not satisfied, then 
the oscillations continue indefinitely as in the top panel of 
Fig.~\ref{phase_3}. Depending on parameters, the scalar field energy density 
may be 
either increasing or decreasing relative to the fluid energy density, 
e.g.~for the 
choices in the figure, $\rho_{{\rm rad}} \propto 1/R^4$ while $\rho_\phi 
\propto  1/R^{4.8}$, so the scalar field becomes less and less 
important.

What happens for $0 < \alpha \le 2$?  For $\alpha = 1$ or $2$ it is easy to
find the exact solutions.  First consider $\alpha = 1$.
Then Eq.~(\ref{phiev3}) becomes
\begin{equation}
\ddot{\phi}  +   {6 \over m} \, {1 \over t} \, \dot{\phi} + 1 = 0 \,,
\end{equation}
and the exact solution is
\begin{equation}
\phi = A + B t^{1-6/m} - {1 \over 2} \, {m \over m+6} \, t^2 \,,
\end{equation}
where $A$ and $B$ are constants to be determined by initial
conditions.  This solution does not display scaling behavior,
and, not surprisingly, $\phi \rightarrow -\infty$ as
$t \rightarrow \infty$, so it is of little physical interest.

The $m=2$ case is more interesting.  For this case, we get the linear 
equation
\begin{equation}
\ddot{\phi} + {6 \over m} \, {1 \over t} \, \dot{\phi} + \phi = 0 \,.
\end{equation}
Taking $\phi = \theta t^{1/2 - 3/m}$, this equation
reduces to
\begin{equation}
t^2 \ddot \theta + t \dot \theta + t^2 \left[1 - (1/2 - 3 / m)^2 \right]
	\theta = 0 \,.
\end{equation}
This is Bessel's equation of order $|1/2 - 3/m|$,
so the general solution for $\phi$ is
\begin{equation}
\label{bessel}
\phi = t^{(1/2 - 3/m)}\left[A J_\nu(t) + B N_\nu(t)\right] \,,
\end{equation}
where $A$ and $B$ are constants determined by the initial
conditions, and $J_\nu$ and $N_\nu$ are Bessel
functions of order $\nu$, with $\nu = |1/2 - 3/m|$.  In the limit of large
$t$, the solutions in Eq.~(\ref{bessel}) all oscillate sinusoidally, with
amplitude decaying as $t^{-3/m}$, so $\rho_\phi \propto R^{-3}$.  Thus, the
solutions for this potential always oscillate, and the density scales as in
Eq.~(\ref{mturner}).

Finally, some exact solutions exist for the case $m = 3$ (matter domination),
for which Eq.~(\ref{phiev3}) reduces to the Lane--Emden equation.
This equation can be solved exactly \cite{Chandra} for $\alpha = 1,2$ and 6,
with the Lane--Emden boundary conditions corresponding to the initial 
condition $\dot
\phi = 0$ at $t=0$.  The solutions for $\alpha=1$ and $\alpha=2$
are special cases
of the solutions discussed above.  The case $\alpha = 6$ corresponds to the
transition between attractor and non-attractor behavior and represents the
analogue (for the matter-dominated case) of the $\alpha = 10$ potential for
the radiation-dominated universe shown in Fig.~\ref{phase_3}.  The phase
diagram in this case resembles the middle diagram in Fig.~\ref{phase_3}.
However, the Lane--Emden solution is {\it not} the $u = \pm 1$ attractor
solution; rather it corresponds to the unstable singular point at $u=0$.
This arises because the Lane--Emden boundary conditions correspond to initial
conditions which lie exactly on the singular point $u=0$, $p=0$, and the 
solution remains there as $t \rightarrow \infty$. The Lane--Emden boundary 
conditions are unphysical when applied to the scalar field evolution 
equation, since $t=0$ is undefined in the cosmological context.

\subsection{The ZWS potential}

Zlatev et al.~\cite{ZWS} made a detailed analysis of the rather unusual 
potential
\begin{equation}
\label{ZWSpot}
V(\phi) \propto \exp \left( \frac{m_{{\rm Pl}}}{\phi} \right) - 1 \,.
\end{equation}
This potential is introduced in recognition of the fact that the simple power
laws do not exhibit the ideal cosmological behavior, in that the scalar field
density grows relative to matter during the matter-dominated era only if it
also grows relative to radiation during the radiation-dominated era.  While
the fine-tuning problem of why the cosmological constant took so long to
dominate is certainly less severe with these power-law potentials than with a
pure $\Lambda$ term, it still remains and one requires either an extremely
low density in the scalar field at early times, or alternatively to have $n$
extremely close to $m$ (i.e.~very large positive or negative $\alpha$) so
that the scalar field requires a very long time to catch up with the
conventional matter, say from an initial state of equipartition with a large
number of fluid components.  Further, the latter resolution, while
superficially attractive, will fail for the same reasons that the exponential
potential does, namely nucleosynthesis and structure formation; its dynamics
are extremely close to the exponential case.

The purpose of a potential such as in Eq.~(\ref{ZWSpot}) is to change the
slope of the scalar field, and hence alter the character of the scaling
solutions with epoch.  For $m_{{\rm Pl}}/\phi \ll 1$, the potential decreases
more rapidly than any power law.  This initial steepness guarantees $n \simeq
m$, and the field is drawn to this approximate tracker behavior.  Later, when
$\phi \gg m_{{\rm Pl}}$, this potential asymptotically approaches the form 
$V(\phi) \propto
1/\phi$, with the scaling solution $n = m/3$, producing scalar
field domination.  This strategy certainly does yield the attractive
observational consequences explored by Zlatev et al.~\cite{ZWS}.  The 
drawback
is that the change of behavior is now governed by the form of the potential,
and not primarily by the equation of state of the accompanying fluid.  That
the scalar field begins to change its behavior around the epoch of
matter--radiation equality is because the feature of changing steepness in
the potential has been placed in the appropriate place.  This represents
tuning of a different sort to the usual tuning of $\Lambda$ models, but a
tuning nonetheless.

\section{Discussion}

Our results indicate that exact solutions for the scalar
field, which give scaling behavior
when the expansion of the universe is driven by a dominant component
with density $\rho_{{\rm dominant}}$,
are possible
for only three classes of potentials:
\begin{enumerate}
\item Exponential potentials ($\rho_\phi$ scales as $\rho_{{\rm dominant}}$).
\item Negative power-law potentials ($\rho_\phi$ decreases less rapidly
than $\rho_{{\rm dominant}}$).
\item Positive power-law potentials ($\rho_\phi$ decreases more rapidly
than $\rho_{{\rm dominant}}$).
\end{enumerate}
The first two cases have been extensively discussed elsewhere;
the existence of the third class is our major new result.
The negative power-law potentials $V \propto \phi^\alpha$
have attractor solutions for all values of the exponent $\alpha$, while
the positive power-law solutions require an exponent $\alpha > 2(6+m)/(6-m)$
for attractor behavior to occur.

Our results do have one practical limitation:  we have confined our attention
to exact solutions.  It is certainly possible, for example, for approximate
solutions to exist which are very close to scaling behavior, e.g., $\rho_\phi
\propto R^{-n} f(R)$, where $f(R)$ is a slowly-varying function of $R$.  If
$f(R)$ varies sufficiently slowly, then there may be no practical distinction
between a solution of this type and our exact solutions.  The particular
potential of Zlatev et al.~is of this type.  It is not practical to
systematically classify all such approximate scaling solutions, although it
may be possible to provide conditions on $V(\phi)$ which allow for such
solutions \cite{Wang}.

%%%%%%%%%%%%%%%%%%%%%%%%%%%%%%%%%%%%%%%%%%%%%%%%%%%%%%%%%%%%%%%%%%%%%%%%
\section*{Acknowledgments}

A.R.L.~was supported by the Royal Society, and R.J.S.~by NASA (NAG 5-2788) at
Fermilab and by the DOE at Fermilab and at Ohio State (DE-FG02-91ER40690).
We thank Limin Wang for useful discussions.  We thank the Fermilab
astrophysics group for hospitality during the start of this work, and
A.R.L.~also thanks the Universit\`{a} di Padova.  A.R.L~acknowledges use of
the Starlink computer system at the University of Sussex.

%%%%%%%%%%%%%%%%%%%%%%%%%%%%%%%%%%%%%%%%%%%%%%%%%%%%%%%%%%%%%%%%%%%%%%%%
 
\end{document}